\documentclass[pra,aps,twocolumn]{revtex4}

\usepackage{amssymb}
\usepackage{amsmath}
\usepackage{epsfig}

\begin{document}
\title{Multiqubit entanglement witness}
\author{Lin Chen}
\author{Yi-Xin Chen}
\affiliation{Zhejiang Insitute of Modern Physics, Zhejiang
University, Hangzhou 310027, China}

\begin{abstract}
We introduce a feasible method of constructing the entanglement
witness that detects the genuine entanglement of a given pure
multiqubit state. We illustrate our method in the scenario of
constructing the witnesses for the multiqubit states that are
broadly theoretically and experimentally investigated. It is shown
that our method can construct the effective witnesses for
experiments. We also investigate the entanglement detection of
symmetric states and mixed states.
\end{abstract}
\maketitle

\section{Introduction}
Quantum nonlocality is one of the most remarkable features
distinguishing the quantum and classical world \cite{Einstein}.
The ``true" nonlocality predicts the existence of entanglement,
which has been proved an extensively useful quantum resource in
quantum information theory (QIT) (for a review see
\cite{DiVincenzo,Horodecki1}). Here we emphasize the word ``true"
in the sense that it is not a kind of classical correlation, and
one cannot prepare it through classical simulation \cite{Werner}.
Quantum nonlocality has been verified in some recent experiments
\cite{Bouwmeester1}.

Motivated by the further understanding of quantum nonlocality and
novel quantm-information processing such as quantum cryptography
\cite{Ekert}, we think it is a meaningful and important job to
tell whether a quantum state is entangled or separable
(classically correlated). Generally, this is difficult, even for
pure states of multipartite system. Several useful theories have
been founded in this context. The Peres-Horodecki criterion gives
a necessary condition on which a bipartite state is separable, and
it is also sufficient for the states in the $2\times2$ or
$2\times3$ Hilbert spaces \cite{Peres}. A sufficient condition
verifying entanglement can be obtained via the violation of Bell
inequality \cite{Bell}. Although the above approaches as well as
the further investigation of them \cite{Cerf,Clauser}, can
effectively detect entanglement of many bipartite states, there
exist some cases in which we need new tools for entanglement
detection \cite{Wolf}. In addition, one will face a more puzzling
situation when applying these theories to multipartite states.

On the other hand, the entanglement witness (EW) was introduced as
a sufficient condition on which one can learn a given state is
entangled or not \cite{Terhal}. The EW is an Hermitian observable,
so it could be used for the experimental demonstration of
entanglement of particular system ( notice the partial transpose
is not completely positive and it cannot be physically realizable
\cite{Nielsen} ). Unlike the Peres-Horodecki criterion and Bell
inequality detecting entanglement in a regular way, one may
construct different types of witnesses for a given state. Adopting
the EW for entanglement detection is thus more flexible than using
the former techniques. By choosing appropriate witness for a
target state, verifying its entanglement via the present
experimental techniques is likely.

In past years, many EWs have been constructed for different
families of entangled states
\cite{Terhal,Doherty,Acin2,Guhne1,Guhne2,Sanpera,Toth}. In
particular, much attentions have been paid to EWs for entangled
multiqubit states due to two main reasons. First, it has been
shown that the multiqubit entanglement lies at the very heart of
quantum-information processing, such as quantum teleportation and
dense coding \cite{Yeo}, error correction \cite{Gottesman},
quantum telecloning \cite{Murao} and quantum computation
\cite{Briegel,Raussendorf}. Second, Many kinds of multiqubit
entangled states have been realized in recent experiments, like
the Greenberger-Horne-Zeilinger (GHZ)
\cite{Bouwmeester2,Resch,Leibfried,Lu}, W
\cite{Eibl,Mikami,Haffner}, cluster \cite{Lu,Walther} and some
special multipartite states \cite{Gaertner,Kiesel2}, by using of
spontaneous parametric down-conversion (SPDC) \cite{Kwiat} and
tomography \cite{James}. The theoretical and experimental progress
indicate that the multiqubit state could be an applicable and
promising quantum resource for the novel tasks in QIT. In this
case, it is important to give a further investigation of EWs for
multiqubit entanglement.

In this paper, we propose a feasible method of constructing the EW
that detects a pure genuine entangled multiqubit state
$\left|\psi\right\rangle_N$ of $N$ parties, i.e., the state whose
reduced density operator of any subsystem has the rank larger than
1. We do it by showing that the state $\left|\psi\right\rangle_N$
can be converted into a state of standard form through some local
invertible local operators (ILOs) \cite{Dur},
$A_0\otimes\cdots\otimes A_{N-1}$ with each nonsingular operator
$A_k,k=0,...,N-1$ acting on the corresponding subsystem of
$\left|\psi\right\rangle_N$. Then, it is easy to construct the EW
for the state of standard form, and we can obtain the witness for
the original state $\left|\psi\right\rangle_N$ via the operators
$A_0\otimes\cdots\otimes A_{N-1}$. Our method is generally
applicable to the multiqubit state of genuine entanglement, and it
does not require the full knowledge of some states. Furthermore,
the proposed EW here for any state $\left|\psi\right\rangle_N$ can
be measured by at most $N^2-N+1$ local devices in experiment.
Compare to the exising method requiring an exponentially
increasing number of measuring devices with $N$ \cite{Sanpera},
our method essentially reduces the necessary experimental effort.
We then respectively construct the EWs for the states which are so
far theoretically and experimentally investigated, including the
two-qubit, three-qubit, GHZ, W, the four-photon state
$\left|\Psi^{(4)}\right\rangle$ \cite{Gaertner}, the four-photon
cluster state \cite{Walther} and the four-photon Dicke state
\cite{Kiesel2}. All the EWs are applicable to experiment. We also
make a study of symmetric multiqubit state on entanglement
detection. Finally, we apply our method to detect mixed state
entanglement.

Our paper is organized as follows. In Sec. II we develop the
method of constructing the EW detecting a pure genuine entangled
multiqubit state. By using of this method and other techniques, we
present the EWs for different states and show their effect when
used in an experiment in Sec. III. In addition, we investigate the
symmetric state and propose some application of our method to
mixed states. We present our conclusion in Sec. IV.

\section{Construction of entanglement witness for multiqubit state}

Let us start by recalling the definition of the EW \cite{Terhal}.
An entanglement witness operator \textbf{W} is an Hermitian
observable which has non-negative expectation values for all
separable states, and thus the entanglement of a particular state
is indicated through the negative expectation value. That is,
\begin{equation}
\mbox{Tr}(\textbf{W}\rho)\geq0, \ \ \rho\in S,
\end{equation}
where $S$ denotes the set of separable states and
\begin{equation}
\mbox{Tr}(\textbf{W}\sigma)<0
\end{equation}
for some genuine entangled state $\sigma$. Since there are other
forms of multipartite states, e.g., the biseparable state, the
tri-separable state and so on \cite{Sanpera}, we use the word
``genuine"  to distinguish them from our target states. We only
consider the pure genuine entangled multiqubit states in this
paper.

A universal EW detecting genuine entanglement close to
$|\psi\rangle$ has been constructed by \cite{Sanpera},
\begin{equation}
\textbf{W}_c= c\ \textbf{I}-\left|\psi\rangle\langle\psi\right|,
c=\underset{|\phi\rangle\in\
B}{\mbox{max}}|\langle\phi|\psi\rangle|^2,
\end{equation}
where $\textbf{I}$ denotes the identity operator and $B$
represents the set of biseparable states. For general $N$-partite
states, determining the value of $c$ is difficult, since one has
to find out the maximal square of the Schmidt coefficient over
$2^{N-1}-1$ possible bipartitions of the state $|\psi\rangle$
\cite{Sanpera}. In addition, the witness $\textbf{W}_c$ often
needs an exponentially increasing number of measuring devices
\cite{Toth}, so more experimental effort is required. Taken in
this sense, constructing the witness $\textbf{W}_c$ is a
universal, but not always applicable method for entanglement
detection.

The concept of ILOs has been firstly introduced to entanglement
manipulation under the criterion of stochastic local operation and
classical communication (SLOCC) \cite{Dur,Thapliyal}. It is the
essential property of the ILOs that an entangled or separable
state will remain entangled or separable after being operated by
some ILOs. The following lemma is thus easily derived from the
property.

\textit{Lemma 1}. Let $\textbf{W}$ be an EW for some $N$-partite
genuine entangled state, and $A_0,...,A_{N-1}$ ILOs. Then
$\textbf{W}^{\prime}=\otimes\prod^{N-1}_{i=0}A_i\textbf{W}\otimes\prod^{N-1}_{i=0}A^{\dag}_i,
$ is also an EW detecting some state of $N$-partite genuine
entanglement. Concretely, if the state $\rho$ is detected by
$\textbf{W}$, then the state
$\rho^{\prime}=\otimes\prod^{N-1}_{i=0}(A^{-1}_i)^{\dag}\rho\otimes\prod^{N-1}_{i=0}A^{-1}_i$
is detected by $\textbf{W}^{\prime}$.
\hspace*{\fill}$\blacksquare$

Such conclusion has been used for entanglement detection recently
\cite{Haffner,Kiesel2}. Clearly, the state $\rho^{\prime}$ is also
completely entangled due to ILOs. Lemma 1 implies that if two
states of genuine entanglement are equivalent under SLOCC, then
one can derive the EW for one of them from the other. So the
existing achievements in entanglement manipulation are helpful to
construct the EW for genuine entanglement. Since what interests us
is the construction of multiqubit witness, a direct way to do it
is to find out all different kinds of multiqubit states under
SLOCC. Despite of a few results in this context
\cite{Dur,Chen1,Chen2,Verstraete}, it is impossible to generally
catalog the multiqubit states due to sophisticated mathematics. In
fact, there is no necessity to find out the classification of
multiqubit states, since we are only concerned about the genuine
entanglement.

For the $N$-partite W state
\begin{equation}
\left|W_N\right\rangle=\frac{1}{\sqrt
N}\Big(\left|1,0,...,0\right\rangle
+\left|0,1,...,0\right\rangle+\cdots+\left|0,0,...,1\right\rangle\Big),
\end{equation}
its witness turns out to be \cite{Sanpera}
\begin{equation}
\textbf{W}_{W_N}=\frac{N-1}{N}\textbf{I}-\left|W_N\rangle\langle
W_N\right|.
\end{equation}
It was first pointed out by D\"ur et al \cite{Dur} that the W
state has a kind of ``robust" entanglement in the sense that the
state remains entangled after losing some particles of the system.
However, we are interested in the witness of more general
entangled state. Let us consider the state
\begin{eqnarray}
\left|\Phi_N\right\rangle&\equiv&a_{1,0}\left|1,0,...,0\right\rangle+a_{1,1}\left|0,1,...,0\right\rangle+\cdots\nonumber\\
&+&a_{1,N-1}\left|0,...,0,1\right\rangle+\sum^{{N\choose2}-1}_{k=0}a_{2,k}P_k(\left|1,1,0,...,0\right\rangle)\nonumber\\
&+&\sum^{{N\choose3}-1}_{k=0}a_{3,k}P_k(\left|1,1,1,0...,0\right\rangle)+\cdots\nonumber\\
&+&a_{N,0}\left|1,1,...,1\right\rangle, a_{1,k}\neq0, k\in[0,N-1],
\end{eqnarray}
where $\{P_k\}$ denotes the set of all distinct permutations of
the spins. This state contains another kind of ``robust"
entanglement in the sense that it is always fully entangled. The
state $\left|\Phi_N\right\rangle$ plays the central role in our
work and we call it the standard multiqubit (SMQ) state. The SMQ
state actually represents a family of multiqubit states, since the
constraint on it is to keep the coefficients $a_{1,k},k=0,...,N-1$
nonvanishing. We have written the terms corresponding to the
coefficients $a_{1,k},k=0,...,N-1$, e.g., $a_{1,0}$'s term is
$\left|1,0,...,0\right\rangle$, $a_{1,1}$'s term is
$\left|0,1,...,0\right\rangle$, etc.

It is easy to show that the SMQ state must be fully entangled. We
rewrite $\left|\Phi_N\right\rangle$ with respect to an arbitrary
bipartition of the system,
\begin{eqnarray}
\left|\Phi_N\right\rangle&=&(a^{\prime}_{1,0}|\overbrace{1,0,...,0}^{m}\rangle+\cdots
+a^{\prime}_{1,m-1}\left|0,...,0,1\right\rangle+\left|X_0\right\rangle)\nonumber\\
&\otimes&|\overbrace{0,0,..,0}^{N-m}\rangle+(a^{\prime}_{1,m}\left|0,0,...,0\right\rangle+\left|X_1\right\rangle)
\otimes|1,0,..,0\rangle\nonumber\\
&+&\left|X_2\right\rangle\otimes|0,1,..,0\rangle+\cdots+\left|X_{2^{N-m}-1}\right\rangle\otimes|1,1,..,1\rangle.
\nonumber\\
\end{eqnarray}
Here, the coefficients $a^{\prime}_{1,k},k=0,...$ come from a
permutation of the initial coefficients $a_{1,k},k=0,...$, due to
the bipartition. Notice the two states $\left|X_0\right\rangle$
and $\left|X_1\right\rangle$ do not contain the term
$\left|0,0,...,0\right\rangle$, which implies the local rank of
the bipartite system is not less than 2. So the SMQ state is
always fully entangled.  \hspace*{\fill}$\blacksquare$

It seems difficult to seek an EW for a generic SMQ state by using
of the existing method in \cite{Sanpera}. We have found a way to
construct the EWs for SMQ states via a special adjustment of the
witness $\textbf{W}_{W_N}$.

\textit{Lemma 2}. Given an SMQ state $\left|\Phi_N\right\rangle$,
it can be detected by a family of witnesses
$\textbf{W}_{SMQ}=\otimes\prod^{N-1}_{k=0}\left(\begin{array}{cc}
1 & 0 \\
0 & \frac{b}{a_{1,k}}
\end{array}\right)^{\dag}\textbf{W}_{W_N}
\otimes\prod^{N-1}_{k=0}\left(\begin{array}{cc}
1 & 0 \\
0 & \frac{b}{a_{1,k}}
\end{array}\right)$, where $b$ is a positive number satisfying the
SMQ-inequality
\begin{eqnarray}
|a_{_{N,0}}a^{-1}_{_{1,0}}a^{-1}_{_{1,1}}\cdots
a^{-1}_{_{1,N-1}}|^2b^{2N-2}+
(|a_{_{N-1,0}}a^{-1}_{_{1,0}}a^{-1}_{_{1,1}}\cdots
a^{-1}_{_{1,N-2}}|^2\nonumber\\
+\cdots+|a_{_{N-1,N-1}}a^{-1}_{_{1,1}}a^{-1}_{_{1,2}}\cdots
a^{-1}_{_{1,N-1}}|^2)b^{2N-4}+\cdots+\nonumber\\
(|a_{_{2,0}}a^{-1}_{_{1,0}}a^{-1}_{_{1,1}}|^2
+\cdots+|a_{_{2,{N\choose2}-1}}a^{-1}_{_{1,N-2}}
a^{-1}_{_{1,N-1}}|^2)b^2<\frac{N}{N-1}.
\end{eqnarray}
Here, $|a_{m,n}|^2$ appears in the term $b^{2m-2}$ in the
polynomial of the left hand side (l.h.s.) of the inequality. The
extra $|a_{1,k}|^{-2}$  is determined by the places of ``1"s in
the term of the state. For example, $a_{_{N,0}}$'s term is
$\left|1,1,...,1\right\rangle$, thus the coefficient of $b^{2N-2}$
is $|a_{_{N,0}}a^{-1}_{_{1,0}}a^{-1}_{_{1,1}}\cdots
a^{-1}_{_{1,N-1}}|^2$; $a_{_{N-1,0}}$'s term is
$\left|1,...,1,0\right\rangle$, thus the coefficient of $b^{2N-4}$
is $|a_{_{N-1,0}}a^{-1}_{_{1,0}}a^{-1}_{_{1,1}}\cdots
a^{-1}_{_{1,N-2}}|^2$, etc.

\textit{Proof}. First, the operator $\textbf{W}_{SMQ}$ is an EW
due to lemma 1. To show it indeed detects the entanglement of
$\left|\Phi_N\right\rangle$, it suffices to show the expectation
value is negative, namely
$\left\langle\Phi_N\right|\textbf{W}_{SMQ}\left|\Phi_N\right\rangle<0$.
This is equivalent to the SMQ-inequality by some simple
calculation.   \hspace*{\fill}$\blacksquare$

One can derive the upper bound $b_{upp}$ of $b$ from this
inequality. An arbitrary choice of $b\in(0,b_{upp})$ leads to an
EW for the given SMQ state. We provide a feasible method for
solving the SMQ-inequality. Suppose the state
$\left|\Phi_N\right\rangle$ is normalized and it is always able to
set $b\leq1$ beforehand. Then the l.h.s. of the SMQ-inequality is
not more than
\begin{eqnarray}
|a^{-1}_{_{1,0}}a^{-1}_{_{1,1}}\cdots
a^{-1}_{_{1,N-1}}|^2(\sum^{{N \choose
i}-1}_{j=0}\sum^{N}_{i=2}|a_{i,j}|^2)b^2=\nonumber\\
\Big[\prod^{N-1}_{i=0}|a_{_{1,i}}|^{-2}(1-\sum^{N-1}_{j=0}|a_{_{1,j}}|^2)\Big]b^2<\frac{N}{N-1}.
\end{eqnarray}
This inequality is easily solvable and the solution of $b$ from it
must make the SMQ-inequality hold. Specially when the SMQ state
has merely the terms
$P_l(\left|1,0,...,0\right\rangle),l=0,...,N-1$, where the
permutation of spins means $P_0(\left|1,0,...,0\right\rangle)=
\left|1,0,...,0\right\rangle,
P_1(\left|1,0,...,0\right\rangle)=\left|0,1,...,0\right\rangle,...,
P_{N-1}(\left|1,0,...,0\right\rangle)=
\left|0,0,...,1\right\rangle,$ the l.h.s. of the SMQ-inequality
equals zero. Hence, any value $b>0$ leads to a witness
$\textbf{W}_{SMQ}$ of this special state. In fact, it is easy to
see that the SMQ state becomes a `pseudo' W state
$\sum_{i=0}^{N-1}a_iP_i(\left|1,0,...,0\right\rangle)$ when all
other terms do not exist, so its witness must be
$\textbf{W}_{SMQ}$ regardless of the change of $b$. It is also
likely to find out other efficient ways to solve the
SMQ-inequality, according to the specific form of given state.

There is another interesting issue we can refer to here. As we
know, it usually requires the full knowledge of the state to be
detected \cite{Horodecki3}. We regard it does hold in the
following context for simplicity. Even so, the SMQ-inequality
indicates we can analytically obtain the upper bound of $b$, if we
are merely aware of the content of coefficients
$a_{1,n},n=0,...,N-1$. So our method can construct the EWs for the
situations in which a little information is provided for the
target SMQ states, which is helpful to the experimental
implementation of multiqubit state by using of tomography
\cite{Resch,Mikami,James}.

Clearly, the set of SMQ states doesn't include all genuine
entangled multiqubit states, e.g., the GHZ state. As mentioned
above, an SMQ state is always fully entangled. In what follows we
consider the inverse question: is every genuine entangled
multiqubit state can be transformed into the SMQ state by some
ILOs? If so, it is then feasible to construct the EWs detecting
genuine entanglement by means of lemma 1 and 2. Suppose the
general multiqubit state is
\begin{equation}
\left|\Psi_N\right\rangle=\sum^{1}_{i_0,i_1,...\atop,i_{N-1}=0}
c_{i_0,i_1,...,i_{N-1}}\left|i_0,i_1,...,i_{N-1}\right\rangle
\end{equation}
and the operator performed on it is
\begin{equation}
V_N=\otimes\prod^{N-1}_{k=0}\left(\begin{array}{cc}
\alpha_{k0} & \alpha_{k1} \\
\alpha_{k2} & \alpha_{k3}
\end{array}\right), \mbox{det}V_N\neq0.
\end{equation}
We have the following result.

\textit{Theorem 1}. Any genuine entangled multiqubit state
$\left|\Psi_N\right\rangle$ can be converted into an SMQ state
$\left|\Phi_N\right\rangle$ by some ILOs $V_N$. The EW detecting
the state $\left|\Psi_N\right\rangle$ is
${V_N}^{\dag}\textbf{W}_{SMQ}V_N$.

\textit{Proof.} See appendix. \hspace*{\fill}$\blacksquare$

So we have given a general method of constructing the EWs
detecting genuine entanglement. If a given multiqubit state is an
SMQ state, one construct its EW by virtue of lemma 2. On the other
hand if a state $\left|\Psi_N\right\rangle$ is not the SMQ state,
one first transform it into an SMQ state by some ILOs $V_N$, whose
witness can be constructed by lemma 1 and lemma 2. In the
appendix, we have provided a method of finding out the ILOs $V_N$
in the first case in terms of $f_k$ and $g_k$, where we set
$\alpha_{k,0}=x^{5^k},\alpha_{k,1}=x^{5^{k+N-1}},k=1,...,N-1$ (
the coefficient of $P_k(\left|1,0,...,0\right\rangle)$ in
$V_N\left|\Psi_N\right\rangle$ can be written as
$\alpha_{k,2}f_k+\alpha_{k,3} g_k,k=0,1,...,N-1,$ see appendix ).
One can thus set the variable $x=e^{i\theta}$, so that the large
exponentials $5^{k},k=1,2,...$ can be removed by the phase. The
weakness of this method is that the finding of $\theta$ which
makes the coefficients of
$P_l(\left|1,0,...,0\right\rangle),l=0,...,N-1$ nonvanishing,
becomes difficult when the number of coefficients
$c_{{i_0},...,i_{N-1}}$ is large. There seems no general method
for this case since the distribution of coefficients is
stochastic, so it is the concrete situation that uniquely
determines how we create the ILOs $V_N$, e.g., when the
coefficients are regularly disposed. Nevertheless, one can try a
tentative method like this. Notice the similarity of coefficients
of $\left|0,0,...,0\right\rangle$ and
$P_l(\left|1,0,...,0\right\rangle),l=0,...,N-1$ in the appendix,
one can set $\alpha_{k,0}=0,\alpha_{k,1}=1,k=1,...,N-1.$ The
criterion of SMQ states then requires
\begin{eqnarray}
c_{0,1,...,1}\alpha_{0,0}+c_{1,1,...,1}\alpha_{0,1}=0,\nonumber\\
c_{0,0,1,...,1}\alpha_{0,0}+c_{1,0,1,...,1}\alpha_{0,1}\neq0,\nonumber\\
\cdots,\nonumber\\
c_{0,1,...,1,0}\alpha_{0,0}+c_{1,1,...,1,0}\alpha_{0,1}\neq0.
\end{eqnarray}
When the coefficients satisfy these relations, one can easily find
out the ILOs $V_N$. If unfortunately, there is four proportional
coefficients such as
$c_{0,1,...,1}/c_{1,1,...,1}=c_{0,0,1,...,1}/c_{1,0,1,...,1},$ the
relations cannot hold. Then one can set
$\alpha_{k,0}=0,\alpha_{k,1}=1,k=0,2,...,N-1,$ and carry out a
similar procedure. In addition, one can set
$\alpha_{k,0}=0,\alpha_{k,1}=1,k=2,3,...,N-1,$ which increases the
number of free variables. It reduces the possibility generating
proportional coefficients. The character of this method is the
necessary amount of calculations is small at the risk of failure.
Generally, it is feasible to construct an EW for the genuine
entangled multiqubit state via our methods.

\section{entanglement detection of practical multiqubit states}

We have proposed a theoretical approach to construct EWs in the
preceding section. Now, let us move to investigate its practical
use for multiqubit states. Two main characters of a practical
witness $\textbf{W}$ in experiments are that, how many measuring
devices are necessary for its realization and how much it
tolerates noise \cite{Terhal}. The EW is an Hermitian operator on
the Hilbert space of $N$ parties. For its experimental
implementation, one has to decompose it into some local von
Neumann measurements $\textbf{W}=\sum^{K}_{k=1}M_k$
\cite{Sanpera}, where each observable $M_k$ is
\begin{eqnarray}
M_k=\sum_{l_0,...,l_{N-1}}b^{(k)}_{l_0,...,l_{N-1}}
|a^{(k,0)}_{l_0}\rangle\langle
a^{(k,0)}_{l_0}|\otimes\cdots\nonumber\\
\otimes|a^{(k,N-1)}_{l_{N-1}}\rangle\langle
a^{(k,N-1)}_{l_{N-1}}|, \ \ \forall b^{(k)}_{l_0,...,l_{N-1}}\in
R.
\end{eqnarray}
The basis $|a^{(k,m)}_{l_m}\rangle$'s are orthogonal vectors for a
fixed $(k,m)$. $m$ is the ordinal number of party, which we will
omit if unnecessary.

It has been shown that each observable $M_k$ can be measured with
one local measuring device in experiments \cite{Terhal}. So when
the number $K$ reaches its minimum, we say the decomposition of
$\textbf{W}$ is optimal. On the other hand, suffering the noise
from the decoherence coupling with the environment is always
unavoidable when implementing quantum-information tasks. The
present experimental techniques require that a superior EW should
be considerably resistant against the noise.

The problem of optimally decomposing a given witness is
technically difficult, and it has been addressed by some authors
\cite{Guhne1,Sanpera,Toth,Guhne2}. For example, the optimal
decomposition of the witness $\textbf{W}_{W_3}$ has been found as
follows \cite{Guhne2}
\begin{eqnarray}
\textbf{W}_{W_3}&=&\frac23\textbf{I}-\left|W_3\rangle\langle W_3\right|\nonumber\\
&=&\frac{1}{24}[(17\cdot\textbf{I}\otimes\textbf{I}\otimes\textbf{I}+
7\cdot\sigma_z\otimes\sigma_z\otimes\sigma_z\nonumber\\
&+&3\cdot\sigma_z\otimes\textbf{I}\otimes\textbf{I}+3\cdot\textbf{I}\otimes\sigma_z\otimes\textbf{I}+
3\cdot\textbf{I}\otimes\textbf{I}\otimes\sigma_z\nonumber\\
&+&5\cdot\sigma_z\otimes\sigma_z\otimes\textbf{I}+5\cdot\textbf{I}\otimes\sigma_z\otimes\sigma_z+
5\cdot\sigma_z\otimes\textbf{I}\otimes\sigma_z)\nonumber\\
&-&(\textbf{I}+\sigma_z+\sigma_x)\otimes(\textbf{I}+\sigma_z+\sigma_x)\otimes(\textbf{I}+\sigma_z+\sigma_x)\nonumber\\
&-&(\textbf{I}+\sigma_z-\sigma_x)\otimes(\textbf{I}+\sigma_z-\sigma_x)\otimes(\textbf{I}+\sigma_z-\sigma_x)\nonumber\\
&-&(\textbf{I}+\sigma_z+\sigma_y)\otimes(\textbf{I}+\sigma_z+\sigma_y)\otimes(\textbf{I}+\sigma_z+\sigma_y)\nonumber\\
&-&(\textbf{I}+\sigma_z-\sigma_y)\otimes(\textbf{I}+\sigma_z-\sigma_y)\otimes(\textbf{I}+\sigma_z-\sigma_y)].\nonumber\\
\end{eqnarray}
According to the definition of decomposition of the witness, 5
local measuring devices are required for the realization of this
witness, namely
$\sigma^{\otimes3}_z,(\textbf{I}+\sigma_z+\sigma_x)^{\otimes3},
(\textbf{I}+\sigma_z-\sigma_x)^{\otimes3},(\textbf{I}+\sigma_z+\sigma_y)^{\otimes3},
(\textbf{I}+\sigma_z-\sigma_y)^{\otimes3}.$ The optimality of the
decomposition has been proved by \cite{Guhne2}. We shall compare
this decomposition with more general result in the present work.

Next, we give a decomposition of the witness $\textbf{W}_{W_N}$.
One can see that there is always a universal way of decomposing it
by using of the identity
$|0,1\rangle\langle1,0|+|1,0\rangle\langle0,1|=\frac12(\sigma_x\sigma_x+\sigma_y\sigma_y).$
That is,
\begin{eqnarray}
\textbf{W}_{W_N}=\frac{N-1}{N}\textbf{I}-\frac{1}{N}\sum_{i=0}^{N-1}
P_i(\left|1,0,...,0\right\rangle)P_i(\left\langle1,0,...,0\right|)\nonumber\\
-\frac{1}{2N}\sum_{i>j=0}^{N-1}(\sigma^{(i)}_x\sigma^{(j)}_x+\sigma^{(i)}_y\sigma^{(j)}_y)\otimes
\prod_{k=0,k\neq i,j}^{N-1}(\left|0\rangle\langle0\right|)^{(k)}.
\end{eqnarray}
The superscripts $(i),i=0,...,N-1$ represent the parties. The
number of measuring settings in this decomposition is $N^2-N+1$,
namely
$\sigma^{\otimes{N}}_z,\sigma^{(i)}_x\sigma^{(j)}_x\prod_{k=0,k\neq
i,j}^{N-1}(\left|0\rangle\langle0\right|)^{(k)},\sigma^{(i)}_y\sigma^{(j)}_y\prod_{k=0,k\neq
i,j}^{N-1}\\
(\left|0\rangle\langle0\right|)^{(k)},i>j=0,...,N-1.$ The
universal decomposition of $\textbf{W}_{W_N}$ requires more
settings than the optimal one, e.g., when $N=3$, it requires $7$,
while the optimal one requires $5$ settings as shown above.
However, such a decomposition is useful to detect the SMQ states.

Recall the form of the witness $\textbf{W}_{SMQ}$ in the lemma 2.
It is easy to see that the entanglement close to an SMQ state
$\left|\Phi_N\right\rangle$ can be detected by using of the
universal decomposition, namely by $N^2-N+1$ measuring devices
\begin{eqnarray}
\sigma^{\otimes{N}}_z,
\Big[\otimes\prod^{N-1}_{k=0}\left(\begin{array}{cc}
1 & 0 \\
0 & \frac{b}{a_{1,k}}
\end{array}\right)^{(k)\dag}\Big]\cdot
\Big[\sigma^{(i)}_x\sigma^{(j)}_x\prod_{k=0,\atop k\neq
i,j}^{N-1}(\left|0\rangle\langle0\right|)^{(k)}\Big]\nonumber\\
\cdot \Big[\otimes\prod^{N-1}_{k=0}\left(\begin{array}{cc}
1 & 0 \\
0 & \frac{b}{a_{1,k}}
\end{array}\right)^{(k)}\Big],
\Big[\otimes\prod^{N-1}_{k=0}\left(\begin{array}{cc}
1 & 0 \\
0 & \frac{b}{a_{1,k}}
\end{array}\right)^{(k)\dag}\Big] \hspace{25pt}\nonumber\\
\cdot\Big[\sigma^{(i)}_y\sigma^{(j)}_y\prod_{k=0,\atop k\neq
i,j}^{N-1}(\left|0\rangle\langle0\right|)^{(k)}\Big]\cdot
\Big[\otimes\prod^{N-1}_{k=0}\left(\begin{array}{cc}
1 & 0 \\
0 & \frac{b}{a_{1,k}}
\end{array}\right)^{(k)}\Big],\hspace{20pt}\nonumber\\
i>j=0,...,N-1. \hspace{153pt}
\end{eqnarray}
The parameter $b$ is determined by the SMQ-inequality. As there
may be better method of decomposing the witness
$\textbf{W}_{SMQ}$, we assert that  one can detect the
entanglement close to an SMQ state through at most $N^2-N+1$
measuring devices. In addition, this result also applies to the
general multiqubit state $\left|\Psi_N\right\rangle$, which is
shown to be converted into some SMQ state
$\left|\Phi_N\right\rangle$ by local unitary operations. According
to theorem 1, there exist some ILOs
$V_N=\otimes\prod^{N-1}_{k=0}\left(\begin{array}{cc}
\alpha_{k0} & \alpha_{k1} \\
\alpha_{k2} & \alpha_{k3}
\end{array}\right)$ such that
$\left|\Phi_N\right\rangle=V_N\left|\Psi_N\right\rangle$. The form
of SMQ state is unchanged under the ILOs
$V^{\prime}_N=\otimes\prod^{N-1}_{k=0}\left(\begin{array}{cc}
x_k & 0 \\
y_k & 1
\end{array}\right),\forall x_k\neq0$. Let
\begin{eqnarray}
x_k=\frac{\alpha_{k1}\alpha_{k2}-\alpha_{k0}\alpha_{k3}}
{|\alpha_{k0}|^2+|\alpha_{k1}|^2},\
y_k=\frac{-\alpha^{*}_{k0}\alpha_{k2}-\alpha^{*}_{k1}\alpha_{k3}}
{|\alpha_{k0}|^2+|\alpha_{k1}|^2}
\end{eqnarray}
for $k=0,1,...,N-1.$ It is easy to check $V^{\prime}_NV_N$ is
unitary after dividing a constant $a$. We can thus transform
$\left|\Psi_N\right\rangle$ into an SMQ state by the unitary
operation $V^{\prime}_NV_N/a$. Because the necessary number of
devices measuring an EW is invariant under the local unitarity,
the following result is hence derived from lemma 1.

\textit{Theorem 2}. An arbitrary genuine entangled multiqubit
state $\left|\Psi_N\right\rangle$ can be transformed into an SMQ
state by local unitary transformation $V_N$. The multiqubit
entanglement close $\left|\Psi_N\right\rangle$ is detected by a
witness $V^{\dag}_N\textbf{W}_{SMQ}V_N$, which can be measured by
at most $N^2-N+1$ local measuring devices $V^{\dag}_NM_kV_N$ with
the settings $M_k,k=1,...,N^2-N+1$ expressed in (16).
\hspace*{\fill}$\blacksquare$

Theorem 2 asserts that we are able to detect the genuine
entanglement of multiqubit state by not more than $N^2-N+1$
devices, which is a polynomial of the party number $N$. Compare to
the existing result in \cite{Sanpera} which often requires an
exponentially increasing number of devices, we have essentially
reduced the necessary experimental effort. One can use the given
method in Sec. II to find out the unitary transformation $V_N$, or
use other tricks based on the specific situation.

To give an example, we consider the experimentally realizable
state \cite{Gaertner,Bourennane2}
\begin{eqnarray}
|\Psi^{(4)}\rangle&=&\frac{1}{\sqrt3}\Big[\left|0011\right\rangle+\left|1100\right\rangle
-\frac12(\left|0110\right\rangle+\left|1001\right\rangle\nonumber\\
&+&\left|0101\right\rangle+\left|1010\right\rangle)\Big],
\end{eqnarray}
whose entanglement has been detected by 15 measuring devices
\cite{Sanpera}. However, our method shows that at most
$4^2-4+1=13$ devices are enough to measure the EW for this state.
Finding the unitary transformation $V_N$ is easy, like $
\left(\begin{array}{cc}
1/\sqrt2 & 1/\sqrt2 \\
1/\sqrt2 & -1/\sqrt2
\end{array}\right)\otimes
\left(\begin{array}{cc}
1/\sqrt5 & 2/\sqrt5 \\
2/\sqrt5 & -1/\sqrt5
\end{array}\right)\otimes
\left(\begin{array}{cc}
3/5 & 4/5 \\
4/5 & -3/5
\end{array}\right)\otimes
\left(\begin{array}{cc}
1 & 0 \\
0 & 1
\end{array}\right)$. Of course one can choose other operation
$V_N$ transforming $\left|\Psi^{(4)}\right\rangle$ into an SMQ
state, for the robustness of witness against the noise is also
considered when the necessary number of devices is unchanged. In
addition, we provide a simple way to reduce the number of
measuring devices. Rewrite the witness $\textbf{W}_{W_4}$ as
follows
\begin{eqnarray}
\textbf{W}_{W_4}&=&\frac34\textbf{I}-\left|W_4\rangle\langle W_4\right|\nonumber\\
&=&\frac34\textbf{I}-\frac14\big[3\left|W_3\rangle\left|0\rangle\langle
W_3\right|\langle0\right|+(\left|0010\rangle\langle0001\right|\nonumber\\
&+&\left|0001\rangle\langle0010\right|)
+(\left|0100\rangle\langle0001\right|+\left|0001\rangle\langle0100\right|)\nonumber\\
&+&(\left|1000\rangle\langle0001\right|+\left|0001\rangle\langle1000\right|)
+\left|0001\rangle\langle0001\right|\big]\nonumber\\.
\end{eqnarray}
We replace the projector $\left|W_3\rangle\langle W_3\right|$ by
$\frac23\textbf{I}-\textbf{W}_{W_3}$, which has the decomposition
requiring 5 devices including $\sigma^{\otimes3}_z$ (see the last
page). The operator in each bracket of equation (19) can be
measured by 2 devices due to the identity
$|0,1\rangle\langle1,0|+|1,0\rangle\langle0,1|=\frac12(\sigma_x\sigma_x+\sigma_y\sigma_y).$
So the witness $\textbf{W}_{W_4}$ has a better decomposition
containing only 11 correlated devices's settings, and so does the
4-body witness $\textbf{W}_{SMQ}$. Since local unitary operations
do not change the necessary number of devices, one can detect the
entanglement of $\left|\Psi^{(4)}\right\rangle$  by 11 devices.
Similarly, one can detect the entanglement of 4-qubit cluster
state \cite{Walther} by using of 11 devices. In what follows we
give more examples to illustrate our techniques.

(i.) The 2-qubit state $\left|\Psi\right\rangle_{AB}$. It is a
kind of state that has been intensively investigated, both
theoretically and experimentally \cite{Nielsen,White}. One can
always write the state as
\begin{eqnarray}
\left|\Psi\right\rangle_{AB}&=&U_A\otimes
U_B(\cos\theta\left|00\right\rangle+\sin\theta\left|11\right\rangle)\nonumber\\
&=&U_A(V_A)^{-1}\otimes U_B(V_B)^{-1}\nonumber\\
&\cdot&[V_A\otimes
V_B(\cos\theta\left|00\right\rangle+\sin\theta\left|11\right\rangle)],
\end{eqnarray}
where unitary operators $U_A$ and $U_B$ are easily known and
$V_A=V_B=\left(\begin{array}{cc}
1 & i\sqrt{\cot\theta} \\
0 & 1
\end{array}\right)$, so the state in the square bracket is an SMQ
state. Then we can find out the unitary transformation $V_N$ based
on the ILOs $U_A(V_A)^{-1}$ and $U_B(V_B)^{-1}$. It implies the
entanglement of every 2-qubit state can be detected via at most
$2^2-2+1=3$ measuring devices. This reaches the same effect as the
witness $(\left|\Psi\rangle_{AB}\langle\Psi\right|)^{T_A}$ in
\cite{Guhne1}. In addition, we investigate the robustness of our
witness against white noise. Let $\theta\in[0,\pi/4]$. The
analytical calculation shows our witness detects a state
$p\textbf{I}/4+(1-p)\left|\Psi\rangle_{AB}\langle\Psi\right|$ with
\begin{equation}
p<\frac{8\cos^2\theta\sin^2\theta}{2-\cos4\theta+\sqrt{2-\cos4\theta-2\sin2\theta}-\sin2\theta}.
\end{equation}
This upper bound is less than
$\frac{2\sin2\theta}{1+2\sin2\theta}$, which is the optimal value
of noise tolerated by the witness
$(\left|\Psi\rangle_{AB}\langle\Psi\right|)^{T_A}$. On the other
hand, the upper bound is better than $\frac43\sin^2\theta$, which
is the optimal value tolerated by the witness $\textbf{W}_c$ in
\cite{Sanpera}.

(ii.) The 3-qubit state $\left|\Psi\right\rangle_{ABC}$. It is
more sophisticated than the 2-qubit state in configuration. There
are two types of genuine entanglement here, namely the GHZ state
and W state. The ILOs converting the 3-qubit states into them can
be found in \cite{Acin1,Dur}. The W state is already an SMQ state,
so it suffices to find out the ILOs converting the GHZ state into
an SMQ state. This is easily done by, e.g.,
$\left(\begin{array}{cc}
1 & -1\\
0 & 1
\end{array}\right)^{\otimes3}
$. By using of these ILOs, we can construct the unitary
transformation $V_N$. As the witness $\textbf{W}_{SMQ}$ can be
measured by 5 devices due to the optimal decomposition of
$\textbf{W}_{W_3}$, we have

\textit{Lemma 3.} The genuine entanglement of every 3-qubit state
can be detected by 5 measuring devices, whose specific forms are
easily obtained in terms of $\textbf{W}_{W_3}$ and the unitary
operator $V_N$. \hspace*{\fill}$\blacksquare$

(iii.) The $N$-partite GHZ state. An optimal witness for this
celebrated state has been constructed in a recent paper
\cite{Toth}, which requires two local measurements. Here we
propose another method of constructing a witness $\textbf{W}$ for
GHZ state. Let $V_N=\left(\begin{array}{cc}
\sqrt2/2 & -\sqrt2/2\cdot(-1)^{\frac{N-1}{N}}\\
\sqrt2/2 & \sqrt2/2\cdot(-1)^{\frac{N-1}{N}}
\end{array}\right)^{\otimes N}$, which transforms the GHZ state
into an SMQ state. So at most $N^2-N+1$ measuring devices are
enough to detect the witness $\textbf{W}_{SMQ}$ for the GHZ state.
Specially, the number of devices can be reduced to 5 when $N=3$ in
view of the optimal decomposition of $\textbf{W}_{W_3}$. This is
better than another usual witness
$\textbf{W}_{1/2}=\frac12\textbf{I}-\left|\mbox{GHZ}\rangle\langle\mbox{GHZ}\right|$,
which is decomposed into $2^N-1$ devices \cite{Toth}. The witness
$\textbf{W}$ can detect a state
$p\textbf{I}/2^N+(1-p)\left|\mbox{GHZ}\rangle\langle\mbox{GHZ}\right|$
with the white noise up to, e.g., $p\approx0.3336$ when $N=3.$ The
ability against the noise gradually declines with an increasing
$N,$ which is similar to the EWs $\textbf{W}_{1/2}$ and that in
\cite{Toth}.

(iv.) The $N$-partite W state. It contains the robust entanglement
\cite{Dur}, and there has been recently remarkable progress in
experimental preparation \cite{Eibl,Mikami,Haffner}. The usual
witness for the $N$-partite W state is $\textbf{W}_{W_N}$ given by
equation (5). We have presented a set of devices measuring the
witness $\textbf{W}_{W_N}$ (see (15) below), which detects a state
$\rho(p)=p\textbf{I}/2^N+(1-p)\left|W\rangle\langle W\right|$ with
white noise up to $p_W<N^{-1}(1-2^{-N})$. So the witness
$\textbf{W}_{W_N}$ becomes weaker and weaker against the noise, as
the number of parties in the system increases. To overcome this
shortcoming, we provide another witness
$\textbf{W}^{\prime}_{W_N}$ via $\textbf{W}_{SMQ}$, namely
\begin{eqnarray}
\textbf{W}^{\prime}_{W_N}=\otimes\prod_{k=0}^{N-1}\left(\begin{array}{cc}
1 & 0\\
0 & b\sqrt N
\end{array}\right)\textbf{W}_{W_N}
\otimes\prod_{k=0}^{N-1}\left(\begin{array}{cc}
1 & 0\\
0 & b\sqrt N
\end{array}\right).
\end{eqnarray}
Then by optimizing Tr$[\rho(p)\textbf{W}^{\prime}_{W_N}]$ with
$b=1/\sqrt{N^2-N}$, one can obtain the upper bound
\begin{equation}
p_{W^{\prime}}<\frac{2^N}{2^N-N+(N-1)^{2-N}N^N},
\end{equation}
which monotonically increases from $N=4.$ The upper bound
$p_{W^{\prime}}>p_W$, and it tends to 1 when $N$ becomes very
large. So the new witness $\textbf{W}^{\prime}_{W_N}$ is more
robust against the noise when the party number increases, which is
different from many existing EWs. On the other hand, the witness
$\textbf{W}^{\prime}_{W_N}$ can be measured by not more than
$N^2-N+1$ devices presented in (16), with
$a_{1,i}=N^{-1/2},i=0,...,N-1$. In particular, the necessary
measuring devices for $\textbf{W}^{\prime}_{W_N}$ seems not more
than that for $\textbf{W}_{W_N}$, e.g., in the case of $N=3$. So
it is more efficient to use the witness
$\textbf{W}^{\prime}_{W_N}$ to detect the W state in experiments.

(v.) The Dicke state $\left|m,N\right\rangle$ \cite{Dicke}. It is
a kind of symmetric state with the form
\begin{equation}
\left|m,N\right\rangle={N \choose
m}^{-1/2}\sum_kP_k(|\overbrace{1,...,1}^{m},0,...,0\rangle).
\end{equation}
The excitation number $m$ ranges from $1$ to $N-1$, and the state
$\left|1,N\right\rangle$ or $\left|N-1,N\right\rangle$ is just the
W state. A theoretical method of detecting entanglement of the
Dicke state around $m=N/2$ is proposed recently, by using of the
witness $\textbf{W}_c$ \cite{Toth3}. There is also the latest
progress in experiment \cite{Kiesel2}, which present the
observation of $\left|2,4\right\rangle$ with a fidelity nearly
$0.844$ by using of quantum tomography. The Dicke state
($m\in[2,N-2]$) also has a kind of robust entanglement in the
sense that it remains entangled if one of the qubit in the state
is projected onto some space. It helps the Dicke state become the
source of EPR singlet, as well as the quantum channel in
telecloning and teleportation \cite{Kiesel2}. So the entanglement
detection of the Dicke state is a valuable work.

One may construct the EW for $\left|2,4\right\rangle$ by using of
the method in \cite{Sanpera} or in the present paper, and the
latter will require a less amount of experimental effort.
Nevertheless, we have found a more efficient way to accomplish the
object. We propose the witness
\begin{eqnarray}
\textbf{W}_{\left|2,4\right\rangle}=
\otimes\prod^3_{k=0}\left(\begin{array}{cc}
1/\sqrt2 & -1/\sqrt2 \\
i/\sqrt2 & i/\sqrt2
\end{array}\right)^{(k)\dag}\Big[2I-\sigma^{\otimes4}_x-\nonumber\\
\frac14\prod_{k=1}^3(\sigma^{(k-1)}_z\sigma^{(k)}_z+I)\Big]\otimes\prod^3_{k=0}\left(\begin{array}{cc}
1/\sqrt2 & -1/\sqrt2 \\
i/\sqrt2 & i/\sqrt2
\end{array}\right)^{(k)}.
\end{eqnarray}
This is indeed an EW since the operator in the square bracket is
the witness $\textbf{W}^{\big( \mbox{GHZ}4 \big)}$ in \cite{Toth},
which can be measured with two measuring devices
$\sigma^{\otimes4}_x$ and $\sigma^{\otimes4}_z$. So it suffices to
measure the witness $\textbf{W}_{\left|2,4\right\rangle}$ by
virtue of only two devices $\Big[\left(\begin{array}{cc}
1/\sqrt2 & -1/\sqrt2 \\
i/\sqrt2 & i/\sqrt2
\end{array}\right)^{\dag}\sigma_x
\left(\begin{array}{cc}
1/\sqrt2 & -1/\sqrt2 \\
i/\sqrt2 & i/\sqrt2
\end{array}\right)\Big]^{\otimes4}$ and
$\Big[\left(\begin{array}{cc}
1/\sqrt2 & -1/\sqrt2 \\
i/\sqrt2 & i/\sqrt2
\end{array}\right)^{\dag}\sigma_z
\left(\begin{array}{cc}
1/\sqrt2 & -1/\sqrt2 \\
i/\sqrt2 & i/\sqrt2
\end{array}\right)\Big]^{\otimes4}$. It is easy to check the
witness $\textbf{W}_{\left|2,4\right\rangle}$ detects entanglement
in the vicinity of state $\left|2,4\right\rangle$, especially it
detects a state
$p\textbf{I}/16+(1-p)\left|2,4\rangle\langle2,4\right|$ with white
noise up to $p<2/9.$

Furthermore, for the generally pure symmetric multiqubit (PSMQ)
state, we investigate some properties of them on entanglement
detection. A PSMQ state is invariant under the exchange of any two
particles in the system \cite{Stockton}. Based on this property,
it is easy to derive that any PSMQ state has the form
\begin{equation}
\left|\Psi_N\right\rangle_{PSMQ}=\sum^{N}_{m=0}c_m\left|m,N\right\rangle,
\end{equation}
where we denote the states
$\left|0,N\right\rangle=\left|0\right\rangle^{\otimes N}$,
$\left|N,N\right\rangle=\left|1\right\rangle^{\otimes N}$.
Evidently, the PSMQ state can be fully entangled such as the Dicke
state, or fully separable such as the state
$\left|0,N\right\rangle$. Then there is a question that whether
the PSMQ state can be partially entangled, namely
$\left|\Psi_N\right\rangle_{PSMQ}=\otimes\prod_{i}\left|\phi_i\right\rangle$
with at least one state $\left|\phi_k\right\rangle$ is fully
entangled. To address it, we extract a part
$\left|\phi_k\right\rangle\otimes\left|\phi_l\right\rangle$ from
the PSMQ state. Suppose there is a particle $A$ in the system $AC$
of $\left|\phi_k\right\rangle$, and $B$ in that of
$\left|\phi_l\right\rangle$. Then, exchanging $A$ and $B$ leads to
the particles $B,C$ are entangled due to the symmetry of the PSMQ
state. It contradicts with the precondition, so the PSMQ state
cannot be partially entangled.

\textit{Lemma 4}. A PSMQ state is either fully entangled or fully
separable. \hspace*{\fill}$\blacksquare$

This character helps construct the EWs for the PSMQ states. One
can divide a PSMQ state into two parts and explore whether this
bipartite state is entangled. Namely a two-body EW is enough to
detect the multiqubit entanglement. On the other hand, since the
PSMQ state is separable if and only if it has the form
$(a\left|0\right\rangle+b\left|1\right\rangle)^{\otimes
N}=a^N\left|0,0,....,0\right\rangle+a^{N-1}b\sum_{l}
P_l(\left|1,0,....,0\right\rangle)+\cdots+b^N\left|1,1,....,1\right\rangle$,
a more efficient method of detecting the symmetric state is to
observe its coefficients $c_m,m=0,...,N.$ This can be implemented
by quantum tomography and only at most $N+1$ coefficients are
necessarily observed. It is also feasible to construct the witness
by using of our method in the appendix, since there are at most
$N+1$ different coefficients in the PSMQ state. This greatly helps
determine the operator $V_N$ that transforms the PSMQ state into
an SMQ state. However, it is not easy to create the witness
$\textbf{W}_c$ for general PSMQ state, though it is highly
symmetric \cite{Toth3}.

It is interesting that one cannot generalize lemma 4 to the case
of mixed symmetric multiqubit (MSMQ) state, whose definition
resembles that of PSMQ state. For example, the 3-body MSMQ state
$\rho_1=\frac13(\left|000\right\rangle+\left|111\right\rangle)
(\left\langle000\right|+\left\langle111\right|)+\frac13\left|111\right\rangle\left\langle111\right|$
is fully entangled since any bipartition of $\rho_1$ is an
entangled state, due to the Peres-Horodecki criterion. On the
other hand, the MSMQ state $\rho_2=\frac12\left|000\right\rangle
\left\langle000\right|+\frac12\left|111\right\rangle\left\langle111\right|$
is fully separable \cite{Werner}. However, we consider the
entangled edge state in expression (14) of \cite{Acin2}
\begin{eqnarray}
\rho_3=\frac{2}{19}\Big[(\left|000\right\rangle+\left|111\right\rangle)
(\left\langle000\right|+\left\langle111\right|)+\nonumber\\
2(\left|001\rangle\langle001\right|+
\left|010\rangle\langle010\right|+
\left|100\rangle\langle100\right|)+\nonumber\\
\frac12(\left|011\rangle\langle011\right|+
\left|101\rangle\langle101\right|+
\left|110\rangle\langle110\right|)\Big],
\end{eqnarray}
which is biseparable with respect to any partition. Because
$\rho_3$ is symmetric, lemma 4 does not hold for the case of MSMQ
states.

The study shows there is a more abundant content of entanglement
detection for mixed states \cite{Barbieri}, so finally we apply
our method to detect the mixed state entanglement. If a given
witness $\textbf{W}$ detects several pure entangled states
$\sigma_i,i=1,2,...$, then it also detects the mixed state
$\sum_{i}a_i\sigma_i,\ a_i\geq0$. By contrast, we often face a
given state $\sigma=\sum_{i}a_i\sigma_i$ to be detected, while it
is difficult to find out an EW detecting each state $\sigma_i$
simultaneously, and thus to detect the state $\sigma$.
Nevertheless, the method of detecting SMQ state sometimes works
here, since one can provide several EWs for the same SMQ state due
to lemma 2. For example, if the state $\sigma_1$ is detected via
$\textbf{W}_{SMQ}$ and $V^{\dag}_N\textbf{W}_{SMQ}V_N$, while the
given state has the decomposition form
$\sigma^{\prime}=\sigma_1+V_N\sigma_1V^{\dag}_N$, then one can
detect the state $\sigma^{\prime}$ by the witness
$\textbf{W}_{SMQ}$. It is feasible to apply this skill to detect
more general state $\rho^{\prime}=\sum_{i}V_i\rho V^{\dag}_i$,
where the state $\rho$ is detected by many EWs
$V^{\dag}_i\textbf{W}_{SMQ}V_i,i=1,...$. The state $\rho^{\prime}$
can be the output of a quantum channel, or more general map taking
$\rho$ as the input state.

In summary, we have found out a decomposition of the witness
$\textbf{W}_{W_N}$. Based on this, it has been shown that not more
than $N^2-N+1$ measuring devices are enough to detect the
entanglement of any $N$-partite SMQ state and furthermore any
$N$-partite genuine entangled multiqubit state
$\left|\Psi_N\right\rangle$. We have explicitly presented the
measuring devices for $\left|\Psi_N\right\rangle$. By using these
facts, we have constructed the EWs for the entanglement detection
of several practical states, including the $|\Psi^{(4)}\rangle$
state, the 2-qubit, 3-qubit, the GHZ, W and the Dicke state. We
also discussed the entanglement detection of PSMQ and MSMQ states,
as well as the mixed states. One can also apply our method to
detect the entanglement of other typical multiqubit states, such
as the GHZ-W-type states \cite{Chen2}.

\section{conclusions}
We have proposed a feasible method of constructing the
entanglement witness detecting the genuine entanglement of pure
multiqubit state. The method is efficient in the sense that it
could reduces the theoretical and experimental effort in many
cases. Our method can be used to construct the entanglement
witnesses for the multiqubit states that are theoretically and
experimentally investigated in literatures. It is a problem that
how to find out the optimal witness for a given state by means of
the proposed techniques, according to the specific requirement of
experiment.

The work was partly supported by the NNSF of China Grant
No.90503009 and 973 Program Grant No.2005CB724508.

\begin{center}
{\bf APPENDIX: CONVERTING A MULTIQUBIT STATE INTO AN SMQ STATE}
\end{center}

Before expanding our proof, we propose a useful proposition.

\textit{Proposition.} Suppose $a>1$ is a natural number,
$\{p_1,p_2,...\}$ is a set of integers with $\forall
p_i\in[1,a-1]$ and similarly for the set $\{q_1,q_2,...\}$.
Suppose two series of natural numbers,
$m_0<m_1<...<m_{k-1},n_0<n_1<...<n_{l-1}$. Then
$\sum^{k-1}_{i=0}p_{m_i}a^{m_i}=\sum^{l-1}_{i=0}q_{n_i}a^{n_i}$ if
and only if $k=l$ and $m_i=n_i,p_{m_i}=q_{m_i},i=0,...,k-1$.

\textit{Proof.} It suffices to verify the necessity. Let
$\sum^{k-1}_{i=0}p_{m_i}a^{m_i}=\sum^{l-1}_{i=0}q_{n_i}a^{n_i}$
and it is no loss of generality to suppose $n_j\geq m_0>n_{j-1}$.
By dividing a factor $a^{m_0}$ on both sides, we have
\begin{equation*}
\sum^{k-1}_{i=0}p_{m_i}a^{m_i-m_0}=\sum^{j-1}_{i=0}q_{n_i}a^{n_i-m_0}+\sum^{l-1}_{i=j}q_{n_i}a^{n_i-m_0}.
\end{equation*}
Clearly, the l.h.s. of the equation is a natural number, and so is
the second term of the r.h.s. of the equation. However, the first
term of the r.h.s. is a proper fraction since
$a^{m_0}-\sum^{j-1}_{i=0}q_{n_i}a^{n_i}\geq
a^{n_{j-1}}-\sum^{j-2}_{i=0}q_{n_i}a^{n_i}\geq\cdots\geq
a^{n_1}-q_{n_0}a^{n_0}>0$, which contradicts with the equation. So
it is only possible that $m_0=n_0$, and hence $p_{m_0}=q_{m_0}$.
Then one obtains a new equation
$\sum^{k-1}_{i=1}p_{m_i}a^{m_i}=\sum^{l-1}_{i=1}q_{n_i}a^{n_i}$.
Repeating the above procedure leads to $m_1=n_1$, and hence
$p_{m_1}=q_{m_1}$. In the same vein one finally verifies the
assertion in the proposition.  \hspace*{\fill}$\blacksquare$

The proposition indeed asserts that the decomposition of a natural
number with respect to some less natural number is unique. This is
similar to the case of binary system, i.e., $N=\sum_i
a_i\cdot2^i,a_i=0$ or 1 is of unique decomposition.

From now on we address the problem of converting a multiqubit
state $\left|\Psi_N\right\rangle$ into an SMQ state. The state
equivalent to $\left|\Psi_N\right\rangle$ under SLOCC is
$V_N\left|\Psi_N\right\rangle$. To find out its relationship with
the SMQ state, we write out some coefficients of terms in
$V_N\left|\Psi_N\right\rangle$,
\begin{eqnarray*}
\left|0,0,...,0\right\rangle: \left(\begin{array}{c}
c_{0,0,...,0,0}  \\
c_{0,0,...,0,1} \\
\cdots  \\
c_{1,1,...,1,1}
\end{array}\right)^T.\Big[\otimes\prod^{N-1}_{k=0}\left(\begin{array}{c}
\alpha_{k,0}  \\
\alpha_{k,1}
\end{array}\right)\Big],\nonumber\\
P_l(\left|1,0,...,0\right\rangle): \left(\begin{array}{c}
c_{0,0,...,0,0}  \\
c_{0,0,...,0,1} \\
\cdots  \\
c_{1,1,...,1,1}
\end{array}\right)^T.\Big[
\otimes\prod^{l-1}_{k=0}\left(\begin{array}{c}
\alpha_{k,0}  \\
\alpha_{k,1}
\end{array}\right)
\nonumber\\
\otimes\left(\begin{array}{c}
\alpha_{l,2}  \\
\alpha_{l,3}
\end{array}\right)
\otimes\prod^{N-1}_{k=l+1}\left(\begin{array}{c}
\alpha_{k,0}  \\
\alpha_{k,1}
\end{array}\right)\Big],l=0,...,N-1.
\end{eqnarray*}
Because an SMQ state contains no the term
$\left|0,0,...,0\right\rangle$, we set
\begin{eqnarray*}
\alpha_{0,0}&=&-\left(\begin{array}{c}
c_{1,0,...,0,0}  \\
c_{1,0,...,0,1} \\
\cdots  \\
c_{1,1,...,1,1}
\end{array}\right)^T.\Big[\otimes\prod^{N-1}_{k=1}\left(\begin{array}{c}
\alpha_{k,0}  \\
\alpha_{k,1}
\end{array}\right)\Big],\nonumber\\
\alpha_{0,1}&=&\left(\begin{array}{c}
c_{0,0,...,0,0}  \\
c_{0,0,...,0,1} \\
\cdots  \\
c_{0,1,...,1,1}
\end{array}\right)^T.\Big[\otimes\prod^{N-1}_{k=1}\left(\begin{array}{c}
\alpha_{k,0}  \\
\alpha_{k,1}
\end{array}\right)\Big].
\end{eqnarray*}
Hence, the coefficient of $P_k(\left|1,0,...,0\right\rangle)$ can
be written as
$\alpha_{k,2}f_k(\alpha_{1,0},\alpha_{1,1},...,\alpha_{N-1,0},\alpha_{N-1,1})+\alpha_{k,3}
g_k(\alpha_{1,0},\alpha_{1,1},...,\alpha_{N-1,0},\alpha_{N-1,1}),k=0,1,...,N-1.$
The equations $f_k$ and $g_k$ are the polynomials of the variables
$\alpha_{i,j},i=1,...,N-1,j=0,1$. For example when $N=2$, it holds
that $f_k=\sum^{2}_{i,j=0}\beta_{i,j}\alpha^i_{1,0}\alpha^j_{1,1}$
with some constant coefficient $\beta_{i,j}$. The SMQ state
requires that every coefficient of
$P_k(\left|1,0,...,0\right\rangle)$ is nonvanishing. Because
$\alpha_{k,2},\alpha_{k,3}$ can be freely determined, we analyze
the situation in terms of $f_k$ and $g_k.$

For the first case, there is no pair of equations $f_k$ and $g_k$
simultaneously identical to zero, namely
$\prod_{k_0,k_1}f_{k_0}g_{k_1}=\sum^{4}_{i_1,...,i_{2N-2}=0}
b_{i_1,...,i_{2N-2}}\prod^{N-1}_{j=1}\prod^{1}_{k=0}\alpha^{i_{j+k(N-1)}}_{j,k}\not\equiv0,k_0\in
S_0, k_1\in S_1,S_0\cup S_1=\{0,1,...,N-1\}$. This implies not
every constant coefficient $b_{i_1,...,i_{2N-2}}$ equals zero. One
can choose nonzero variables $\alpha_{i,j},i=1,...,N-1,j=0,1$
making the product $\prod_{k_0,k_1}f_{k_0}g_{k_1}\neq0$.  For
example, let
$\alpha_{k,0}=x^{5^k},\alpha_{k,1}=x^{5^{k+N-1}},k=1,...,N-1$, $x$
is a variable. Then the powers of $x$,
$\prod^{N-1}_{j=1}\prod^{1}_{k=0}\alpha^{i_{j+k(N-1)}}_{j,k},\forall
i_j=0,1,2,3,4$ have different exponentials in terms of the
proposition. So the polynomial equation
$\prod_{k_0,k_1}f_{k_0}g_{k_1}(x)=0$ has a finite number of
solutions $x\in S_2$. Similar results are applicable to
$\alpha_{0,0}$ and $\alpha_{0,1}$, which have finite solution sets
$S_3$ and $S_4$, respectively. Then we choose $x\not\in S_2\cup
S_3\cup S_4$, and suitable $\alpha_{k,2},\alpha_{k,3},k=0,...,N-1$
making every coefficient of $P_k(\left|1,0,...,0\right\rangle)$
nonvanishing, as well as the non-singularity of the operator
$V_N$. So we have transformed the initial state into an SMQ state.

For the second case, there is at least a pair of equations $f_k$
and $g_k$ simultaneously identical to zero no matter how the
variables $\alpha_{k,0},\alpha_{k,1},k=1,...,N-1$ change, which
means at least one term $P_k(\left|1,0,...,0\right\rangle)$ is
always removed. We show in this case that the state
$\left|\Psi_N\right\rangle$ must be separable. For the case of
$f_0=g_0\equiv0$ namely $\alpha_{0,0}=\alpha_{0,1}\equiv0$,
choosing
$\alpha_{k,0}=x^{5^k},\alpha_{k,1}=x^{5^{k+N-1}},k=1,...,N-1$
leads to that every coefficient $c_{i_0,i_1,...,i_{N-1}}$ equals
zero, due to the proposition. So it is impossible that
$f_0=g_0\equiv0$.

On the other hand, there may be the case of
$f_k=g_k\equiv0,k\in[1,N-1]$. It suffices to investigate the case
of $f_1=g_1\equiv0,$ and other situations can be similarly dealt
with. Taking into account the expressions of $f_1$ and $g_1$ and
removing the parameters $\alpha_{0,0},\alpha_{0,1}$, we have
\begin{equation*}
\frac{\vec{c}_{1,0}.\Big[\otimes\prod^{N-1}_{k=2}\left(\begin{array}{c}
\alpha_{k,0}  \\
\alpha_{k,1}
\end{array}\right)\Big]}{\vec{c}_{0,0}
.\Big[\otimes\prod^{N-1}_{k=2}\left(\begin{array}{c}
\alpha_{k,0}  \\
\alpha_{k,1}
\end{array}\right)\Big]}=
\frac{\vec{c}_{1,1}.\Big[\otimes\prod^{N-1}_{k=2}\left(\begin{array}{c}
\alpha_{k,0}  \\
\alpha_{k,1}
\end{array}\right)\Big]}{\vec{c}_{0,1}
.\Big[\otimes\prod^{N-1}_{k=2}\left(\begin{array}{c}
\alpha_{k,0}  \\
\alpha_{k,1}
\end{array}\right)\Big]},(*)
\end{equation*}
where the $2^{N-2}\times1$ coefficient vector $\vec{c}_{i,j}$
represents
\begin{eqnarray*}
\vec{c}_{i,j}=\left(\begin{array}{c}
c_{i,j,0,...,0,0}  \\
c_{i,j,0,...,0,1} \\
\cdots  \\
c_{i,j,1,...,1,1}
\end{array}\right)^T,i,j=0,1,
\end{eqnarray*}
and similarly for $\vec{c}_{i,j,k},i,j,k=0,1.$ Let us analyze the
condition making the equation $(*)$ an identity. If some vector
$\vec{c}_{i,j}=0,$ then it holds  that $\vec{c}_{1-i,j}=0$ or
$\vec{c}_{i,1-j}=0$, which is derived by choosing
$\alpha_{k,0}=x^{5^k},\alpha_{k,1}=x^{5^{k+N-1}},k=1,...,N-1$
again. In this case, the state $\left|\Psi_N\right\rangle$ is
separable. In what follows we suppose no vector
$\vec{c}_{i,j}=0,i,j=0,1.$

For the case of $N=3,$ the equation $(*)$ becomes
\begin{equation*}
\frac{c_{1,0,0}\alpha_{2,0}+c_{1,0,1}\alpha_{2,1}}{c_{0,0,0}\alpha_{2,0}+c_{0,0,1}\alpha_{2,1}}=
\frac{c_{1,1,0}\alpha_{2,0}+c_{1,1,1}\alpha_{2,1}}{c_{0,1,0}\alpha_{2,0}+c_{0,1,1}\alpha_{2,1}}.
\end{equation*}
As the variables $\alpha_{2,0}$ and $\alpha_{2,1}$ arbitrarily
change, simple algebra leads to
$\vec{c}_{1,0}=k_1\vec{c}_{0,0},\vec{c}_{1,1}=k_1\vec{c}_{0,1},$
or
$\vec{c}_{1,0}=k_2\vec{c}_{1,1},\vec{c}_{0,0}=k_2\vec{c}_{0,1},$
with $k_1,k_2$ two proportional constants. Either of them makes
$(*)$ an identity. Suppose the result applies to the case of
$N=m$, namely if the identity $(*)$ holds then it always holds
$\vec{c}_{1,0}=k_1\vec{c}_{0,0},\vec{c}_{1,1}=k_1\vec{c}_{0,1},$
or
$\vec{c}_{1,0}=k_2\vec{c}_{1,1},\vec{c}_{0,0}=k_2\vec{c}_{0,1}$.
For the case of $N=m+1$, we rewrite the equation $(*)$ as
\begin{eqnarray*}
\frac{(\vec{c}_{1,0,0}\alpha_{2,0}+\vec{c}_{1,0,1}\alpha_{2,1}).\Big[\otimes\prod^{m}_{k=3}\left(\begin{array}{c}
\alpha_{k,0}  \\
\alpha_{k,1}
\end{array}\right)\Big]}
{(\vec{c}_{0,0,0}\alpha_{2,0}+\vec{c}_{0,0,1}\alpha_{2,1}).\Big[\otimes\prod^{m}_{k=3}\left(\begin{array}{c}
\alpha_{k,0}  \\
\alpha_{k,1}
\end{array}\right)\Big]}=\nonumber\\
\frac{(\vec{c}_{1,1,0}\alpha_{2,0}+\vec{c}_{1,1,1}\alpha_{2,1}).\Big[\otimes\prod^{m}_{k=3}\left(\begin{array}{c}
\alpha_{k,0}  \\
\alpha_{k,1}
\end{array}\right)\Big]}
{(\vec{c}_{0,1,0}\alpha_{2,0}+\vec{c}_{0,1,1}\alpha_{2,1}).\Big[\otimes\prod^{m}_{k=3}\left(\begin{array}{c}
\alpha_{k,0}  \\
\alpha_{k,1}
\end{array}\right)\Big]}.
\end{eqnarray*}
Applying the assumption that the case of $N=m$ holds to this
expression, we have
\begin{eqnarray*}
\vec{c}_{1,0,0}\alpha_{2,0}+\vec{c}_{1,0,1}\alpha_{2,1}=
k(\vec{c}_{0,0,0}\alpha_{2,0}+\vec{c}_{0,0,1}\alpha_{2,1}), \ \ \  (*.1)\\
\vec{c}_{1,1,0}\alpha_{2,0}+\vec{c}_{1,1,1}\alpha_{2,1}=
k(\vec{c}_{0,1,0}\alpha_{2,0}+\vec{c}_{0,1,1}\alpha_{2,1}), \ \ \
(*.2)
\end{eqnarray*}
or
\begin{eqnarray*}
\vec{c}_{1,0,0}\alpha_{2,0}+\vec{c}_{1,0,1}\alpha_{2,1}=
k(\vec{c}_{1,1,0}\alpha_{2,0}+\vec{c}_{1,1,1}\alpha_{2,1}),\ \ \ (*.3)\\
\vec{c}_{0,0,0}\alpha_{2,0}+\vec{c}_{0,0,1}\alpha_{2,1}=
k(\vec{c}_{0,1,0}\alpha_{2,0}+\vec{c}_{0,1,1}\alpha_{2,1}).\ \ \
(*.4)
\end{eqnarray*}
We first analyze equations $(*.1)$ and $(*.2)$, in which the
proportional number $k$ can be the function of $\alpha_{2,0}$ and
$\alpha_{2,1}$. Let the constant $C^{l}_{i,j,k}$ be the $l$'th
entry of the $2^{m-2}\times1$ vector $\vec{c}_{i,j,k},i,j,k=0,1.$
Similar to the case of $N=3$, we have
\begin{equation*}
\frac{C^{l_1}_{1,0,0}}{C^{l_1}_{0,0,0}}=\frac{C^{l_1}_{1,0,1}}{C^{l_1}_{0,0,1}}
=\frac{C^{l_1}_{1,1,0}}{C^{l_1}_{0,1,0}}=\frac{C^{l_1}_{1,1,1}}{C^{l_1}_{0,1,1}}
\ \ \ \ \ (*.5)
\end{equation*}
or
\begin{equation*}
\frac{C^{l_2}_{1,0,0}}{C^{l_2}_{1,1,0}}=\frac{C^{l_2}_{1,0,1}}{C^{l_2}_{1,1,1}}
=\frac{C^{l_2}_{0,0,0}}{C^{l_2}_{0,1,0}}=\frac{C^{l_2}_{0,0,1}}{C^{l_2}_{0,1,1}}
\ \ \ \ \ (*.6)
\end{equation*}
with $l_1\in S_5,l_2\in S_6,S_5\cup S_6=\{0,1,...,2^{m-2}-1\}$. If
the equation $(*.5)$ holds for some $l_1$, it is easy to verify
$k$ is a constant and thus
$\vec{c}_{1,0}=k\vec{c}_{0,0},\vec{c}_{1,1}=k\vec{c}_{0,1}.$ If no
equation $(*.5)$ holds, namely the equation $(*.6)$ holds for any
$l_2\in[0,2^{m-2}-1]$, we have
$\vec{c}_{1,0}=k^{\prime}\vec{c}_{1,1},\vec{c}_{0,0}=k^{\prime}\vec{c}_{0,1}$.
Analyzing the equations $(*.3)$ and $(*.4)$ leads to the same
conclusion. So we have shown by induction that the equation $(*)$
becomes an identity if and only if
$\vec{c}_{1,0}=k_1\vec{c}_{0,0},\vec{c}_{1,1}=k_1\vec{c}_{0,1},$
or
$\vec{c}_{1,0}=k_2\vec{c}_{1,1},\vec{c}_{0,0}=k_2\vec{c}_{0,1}$.
Either of them asserts the state $\left|\Psi_N\right\rangle$ is
separable, concretely
$\left|\Psi_N\right\rangle=\left|\psi_0\right\rangle\otimes\left|\psi_{1,2,...,N-1}\right\rangle$
or
$\left|\Psi_N\right\rangle=\left|\psi_1\right\rangle\otimes\left|\psi_{0,2,...,N-1}\right\rangle$,
respectively. This completes the proof for $f_1=g_1\equiv0.$

For other cases $f_k=g_k\equiv0,k=2,...,N-1$, it holds that the
state
$\left|\Psi_N\right\rangle=\left|\psi_0\right\rangle\otimes\left|\psi_{1,2,...,N-1}\right\rangle$
or
$\left|\Psi_N\right\rangle=\left|\psi_k\right\rangle\otimes\left|\psi_{0,...,k-1,k+1,...,N-1}\right\rangle$,
which can be verified by following the technique similar to the
proof for $f_1=g_1\equiv0.$ In conclusion, if a multiqubit state
$\left|\Psi_N\right\rangle$ cannot be converted into an SMQ state
via some ILOs, then it must be separable. It means the genuine
entangled multiqubit state can always be converted into an SMQ
state via some ILOs.

\end{document}